\documentclass{nature}
\usepackage{graphicx}
\usepackage{amssymb}
\usepackage{amsfonts}
\usepackage{caption}
\usepackage{bm}
\usepackage{epstopdf}
\usepackage{epsfig}
\begin{document}

\title{Engineering Topological Surface States and
Giant Rashba Spin Splitting in BiTeI/Bi$_2$Te$_3$ Heterostructures}

\author{Jin-Jian Zhou$^{1}$, Wanxiang Feng$^{2}$, Ying Zhang$^{4}$, Shengyuan A. Yang$^{3,\star}$, Yugui Yao$^{2,\star}$}

\maketitle

\begin{affiliations}
\item Institute of Physics, Chinese Academy of Sciences and Beijing
National Laboratory for Condensed Matter Physics, Beijing 100190, China
\item School of Physics, Beijing Institute of Technology, Beijing 100081, China
\item Engineering Product Development, Singapore University of Technology and Design, Singapore 138682, Singapore
\item Department of Physics, Beijing Normal University,
Beijing 100875, China

$^\star$e-mail: shengyuan\_yang@sutd.edu.sg; ygyao@bit.edu.cn
\end{affiliations}

\begin{abstract}
  The search for strongly inversion asymmetric topological insulators is an active research field
  because these materials possess distinct properties compared with the inversion symmetric ones. In particular, it is desirable to realize a large Rashba spin-splitting (RSS)
  in such materials, which combined with the topological surface states (TSS) could lead to
  useful spintronics applications. In this report, based on first principles calculations, we predict that the heterostructure of BiTeI/Bi$_{2}$Te$_{3}$ is a strong topological insulator with a giant RSS. The coexistence of TSS and RSS in the current system is native and stable. More importantly, we find that both the $\mathbb{Z}_{2}$ invariants and the Rashba energy can be controlled by engineering the layer geometries of the heterostructure, and the Rashba energy can be made even larger than that of bulk BiTeI.
  Our work opens a new route for designing topological spintronics devices based on inversion asymmetric heterostructures.
\end{abstract}

The study of the effects of spin-orbit coupling (SOC) has been one of the central themes in physics research in the past decade. A significant discovery is that a strong SOC can convert a normal insulator into a topological insulating phase, in which metallic surface states appear in the bulk gap and they are protected by a nontrivial bulk topological order\cite{Hasan2010,Qi2011}. Another important consequence of SOC is the Rashba effect in inversion asymmetric structures\cite{Rashba1960,Bychkov1984}. It has been extensively studied in the context of spintronics, which aims to utilize the electron's spin degree of freedom for device applications. It is of great interest to search for a material hosting both robust TSS and large RSS which could be used as a playground to study the interplay between these two interesting effects and also as a base material for technological applications.

A nonzero RSS requires structural inversion symmetry breaking. 
It occurs naturally at sample surfaces or interfaces. For example, RSS has been observed at metallic surfaces\cite{LaShell1996,Koroteev2004,Ast2007}, ultrathin metal films\cite{Frantzeskakis2008,He2008}, and semiconductor heterostructures\cite{Nitta1997}. In particular, ARPES experiments on the topological insulator Bi$_2$Se$_3$ have shown that RSS exists at surfaces with artificially deposited adsorbates\cite{King2011,Zhu2011,Bianchi2011,Benia2011}. Such RSS, however, are typically small and sensitive to the surface environment. 
On the other hand, a large RSS of the bulk bands is possible in polar compounds with heavy elements, such as the noncentrosymmetric layered semiconductor BiTeI which was recently reported to have the largest known RSS strength\cite{Ishizaka2011}. However, BiTeI is a trivial insulator at ambient condition. It was proposed that under pressure BiTeI might be turned into a topological insulator\cite{Bahramy2012}. But pressure induced structural transition might also occur as suggested by experiments\cite{Tran2013,Xi2013}.

Motivated by the recent
progress in the experimental fabrication of layered heterostructures\cite{Geim2013,Haigh2012}, here we predict that the coexistence of robust TSS and giant RSS can be realized in the [(BiTeI)$_m$(Bi$_2$Te$_3$)$_n$] heterostructure in its native state without any need of strain or pressure ($m$ and $n$ represent
the number of layers of each constituent in a unit cell). Moreover, we find that
both the topological order and the RSS energy can be controlled by varying the number of layers.
In particular, the $\mathbb{Z}_2$ topological invariants of heterostructure with even (odd) $m$ are given by [1;000] ([1;001]). And the RSS strength is tunable by varying the ratio between $m$ and $n$.  Especially the structures [(BiTeI)$_2$(Bi$_2$Te$_3$)$_1$]  and [(BiTeI)$_3$(Bi$_2$Te$_3$)$_1$]  have Rashba energies even larger than that of BiTeI. The TSS at different crystalline surfaces are analyzed in detail, which demonstrate characteristic features of a strongly inversion asymmetric topological insulator. Our findings thus identify a promising topological material for future spintronics and quantum computation applications.

\section*{Results}
\subsection{Basic crystal structure.}
The well known topological insulator material Bi$_2$Te$_3$ has a layered structure with five atomic layers (Te-Bi-Te-Bi-Te) as a
quintuple layer (QL) unit\cite{Zhang2009}. BiTeI has a similar layered structure with three atomic layers (Te-Bi-I) as a triple layer (TL) unit\cite{Ishizaka2011}.  The interactions between QLs (or TLs) are of weak van-der-Waals (vdW) type, while the interactions within one QL (or TL) are of strong covalent or ionic type. Since in-plane lattice constants of Bi$_2$Te$_3$ (4.383~\AA) and BiTeI (4.339~\AA) are very close, their heterostructure [(BiTeI)$_m$(Bi$_2$Te$_3$)$_n$], which consists of alternating $m$ TLs of  BiTeI and $n$ QLs of Bi$_2 $Te$_3 $, is possible to be epitaxially grown along the crystallographic $c$-axis\cite{Isaeva2013}.  The simplest structure [(BiTeI)$_1$(Bi$_2 $Te$_3 $)$_1$] (hereafter referred to as Bi$_3 $Te$_4 $I) is shown in Fig.~1(a). In the following, we shall choose it as a representative example for the discussion of general structural and electronic properties of this family of heterostructures.

The optimized lattice parameters and bond lengths without and with vdW corrections
are listed in Table.I.  As can be seen from the table,  the inter-layer bond lengths (I-Te$_1$ and Te$_3$-Te$_4$) are much larger than the intra-layer bond lengths in both cases. With vdW correction,  the intra-layer bond lengths are hardly changed, while the inter-layer bond lengths are significantly reduced, confirming that the interactions between TL and its adjacent QLs are of weak and non-local vdW type. The Te$_3$-Te$_4$ and I-Te$_1$ bond lengths relaxed by vdW-DF method are similar to the inter-layer Te-Te bond length in Bi$_2$Te$_3$ (3.748~\AA) and the I-Te bond length in BiTeI (3.924~\AA) respectively, which is reasonable. The dynamic stability of the structure is further investigated through the phonon spectrum calculations.  The calculated phonon spectrum with SOC is shown in Fig.~1(c).  The absence of imaginary frequencies demonstrates that our structure of Bi$_3$Te$_4$I is dynamically stable.
In the following electronic calculations, the optimized structure with vdW corrections is adopted.

\subsection{Giant Rashba spin splitting.}
The calculated band structure of Bi$_3$Te$_4$I without SOC is shown in Fig.~2(a).
As we can see, Bi$_3$Te$_4$I is a direct band gap semiconductor with E$_g \approx$ 0.52 eV,  much smaller than
that of BiTeI ($\approx$ 1.2 eV)\cite{Bahramy2011}.
The direct band gap is located at A-point of the Brillouin zone (labeled in Fig.~1(b)). The states near the band edges are dominated by Bi-6$p_{z}$ orbitals
for the conduction band bottom, and by Te-5$p_{z}$ orbitals for the valence band top.

When SOC is turned on, as shown in Fig.~2(b), 
the energy bands are spin-splitted and at the same time the conduction band (local) minima
are shifted away from the high symmetry points $\Gamma$ and A, which is similar to the case of bulk BiTeI.
This is the signature of a Rashba type spin splitting\cite{Ishizaka2011}. The RSS is characterized by the Rashba energy $E_R$, 
the momentum offset $k_0$ and the Rashba coupling parameter $\alpha_R = 2E_R/k_0$. These parameters 
are illustrated in Fig.~2(c) for the conduction band bottom near $\Gamma$ point. The RSS at 
$\Gamma$ point is slightly anisotropic. Along $\Gamma$-K direction, we have $E_R\approx76$ meV and 
$\alpha_R\approx2.54$ eV$\cdot$\AA, while along the $\Gamma$-M direction, the corresponding values are smaller, with $E_R\approx74$ meV and $\alpha_R\approx2.48$ eV$\cdot$\AA. The RSS of conduction band near A point has a slightly smaller Rashba energy ($E_R\approx56$ meV), but the value of $\alpha_R$ ($\approx$ 2.51 eV$\cdot$\AA) is close to that of the $\Gamma$ point.
The value of RSS in the heterostructure Bi$_3$Te$_4$I is much larger than the RSS of conventional semiconductor heterostructures (on the order of  meV). 
This giant RSS can be viewed as being inherited from the BiTeI component, which in its bulk form has a RSS of $113$ meV ( $\alpha_R\approx4.52$ eV$\cdot$\AA )\cite{Bahramy2011}. 

\subsection{Topological insulating phase.}
In addition to RSS, the band order is also inverted after the introduction of SOC.
In the presence of SOC, spin and orbital angular momentum are mixed and the $p_z$ states are transformed into $j=1/2$ states. As a consequence, the conduction band bottom, which is mainly occupied by the Bi-$p_z$ orbital, is pushed downward. Meanwhile the valence band top, which is occupied by the Te-$p_z$ orbital, is pushed upward. Due to the large atomic SOC from Bi and Te, the energy gap around A point is inverted. To facilitate
our understanding of this band inversion process, we gradually increase the strength of SOC ($\lambda_{SO}$) from zero to its full value, and monitor the evolution of the band structure. The local band gap at $k_z$=0.5 plane (around A points) and at $k_z$=0.0 plane (around $\Gamma$ point) is plotted in Fig.~2(d). As $\lambda_{SO}$ increases, the local gap around $\Gamma$ points decreases from 0.85 eV to 0.12 eV without closing. In contrast, the local gap around A points decreases, closes, reopens, and finally reaches a value of 0.22 eV. This band inversion induced by SOC
signals a topological phase transition in the current system. 

To pin down the topological phase of the heterostructure more precisely, 
we calculate the topological $\mathbb{Z}_{2}$ invariants.
Since the inversion symmetry is explicitly broken in Bi$_{3}$Te$_{4}$I, the parity criterion can not apply\cite{Fu2007}. Instead, we evaluate the $\mathbb{Z}_{2}$ invariants $\nu_{0};(\nu_{1}\nu_{2 }\nu_{3})$ in terms of the Berry gauge potential and Berry curvature by using our recently developed general method within the full-potential linearized augmented plane wave framework\cite{Feng2012}.  The calculated $\mathbb{Z}_{2}=1;(001)$ for Bi$_3$Te$_4$I is consistent with the previous band inversion analysis. 
At first sight, this topological insulating phase seems to be inherited from the Bi$_2$Te$_3$ component. However, it should be noted
that the $\mathbb{Z}_{2}$ invariants of Bi$_3$Te$_4$I is different from that of Bi$_2$Te$_3$ ($\mathbb{Z}_{2}=1;(000)$). Instead it is more like 
bulk BiTeI under pressure\cite{Bahramy2012}, which has the same topological invariants. The difference in the index $\nu_3$ reflects the different band inversion point in the Brillouin zone. For Bi$_2$Te$_3$, band
inversion occurs at $\Gamma$ point, while for Bi$_3$Te$_4$I band inversion occurs at A point. The nonzero value of $\nu_{3}$ can have interesting physical consequences. For example, it has been shown that a system with $\nu_{3}=1$ may 
offer a topologically protected one-dimensional conducting channel along the $c$-axis\cite{Ran2009}.

The nontrivial $\mathbb{Z}_2$ value is tied to the existence of TSS\cite{Fu2007a}. As another verification, we calculate the surface energy spectrum directly. Fig.~2(e) shows the local density of states (LDOS) of a side-cut surface (the $ac$ plane of Bi$_3$Te$_4$I). A Dirac cone in the bulk energy gap can be clearly seen, which is located at $\bar{A}$ point of the projected two-dimensional Brillouin zone. This again confirms the topological insulating phase of Bi$_3$Te$_4$I.

\subsection{Controllability of properties.}
Thus far, we have shown that the simplest heterostructure Bi$_3$Te$_4$I  was a strong topological insulator with giant bulk RSS. Now we extend our investigation to the family of heterostructures [(BiTeI)$_m$(Bi$_2$Te$_3$)$_n$] with different number of layers $m$ and $n$, and see how the topological invariants and the RSS depend on the layer configuration. The calculated results are summarized in Table.II.   We observe that the coexistence of strong topological insulating phase and giant bulk RSS is common to all these heterostructures. Quite interestingly, the $\nu_3$ index of $\mathbb{Z}_2$ has a dependence on the parity of $m$. Namely $\nu_3=1$ when $m$ is odd and  $\nu_3=0$ when $m$ is even.  For odd $m$, band inversion occurs at A point, while for even $m$, it occurs at $\Gamma$ point. The location of the surface Dirac cone is associated with the band inversion point, hence can also be controlled by the parity of $m$. (Note that this is meaningful for side surfaces where the projections of A and $\Gamma$ are distinguishable.)  From the results, we see that the Rashba energy is also tunable to a great extent by varying $m$ and $n$.  $E_R$ generally increases with larger $m$ and smaller $n$.
It should be noted that heterostructures [(BiTeI)$_2$(Bi$_2$Te$_3$)$_1$]  and [(BiTeI)$_3$(Bi$_2$Te$_3$)$_1$] have extremely large $E_R$ of 130 meV ($\alpha_R\approx5.23$ eV$\cdot$\AA) and 139 meV ($\alpha_R\approx5.45$ eV$\cdot$\AA) respectively, which are both larger than that of bulk BiTeI ($\approx$ 113 meV) reported before\cite{Bahramy2011}.

\subsection{Surface states.}
A salient feature of an inversion asymmetric topological insulator is that different crystal surfaces have different
TSS. In our heterostructure system, due to the giant RSS and the freedom to control the properties by layer configurations, more interesting
physics are expected to appear on the sample surface. 
As two representative examples with different $\mathbb{Z}_2$ invariants, we shall discuss Bi$_{3}$Te$_{4}$I with $\mathbb{Z}_2=1;(001)$ and Bi$_4$Te$_5$I$_2$ ([(BiTeI)$_2$(Bi$_2$Te$_3$)$_1$]) with $\mathbb{Z}_2=1;(000)$ in the following.

Due to the lack of inversion symmetry, there are two different surfaces of Bi$_3$Te$_4$I (or Bi$_4$Te$_5$I$_2$) perpendicular to the $c$-axis, namely the I-terminated surface and the Te-terminated surface. They are constructed by cleaving the adjacent QL and TL (TL and TL for Bi$_4$Te$_5$I$_2$) through the weakly coupled Te-I gap. As can be seen in Fig.~3(a,c),
although with different $\nu_{3}$ index in $\mathbb{Z}_2$,  the two heterostructures have similar surface spectra. There are two different kinds of electronic states in the energy window around the bulk band gap. The states with dispersions identical for both I-terminated and Te-terminated surfaces are the bulk RSS states whose wavefunctions spread over the bulk region, thus are insensitive to the surface environment. The other states with linear dispersions that cross the bulk gap are the TSS. Owing to the strong inversion asymmetry, TSS are very different at different atom-terminated surfaces. For both heterostructures, the Dirac point of TSS stays in bulk band gap at I-terminated surface [Fig.~3(a, c)],
while it is buried in bulk valence bands at Te-terminated surface [Fig.~3(b, d)]. If Fermi level lies in the bulk gap below the Dirac point 
of I-terminated surface, the two surfaces of a sample would have different types of charge carriers, which can be used to construct a topological
p-n junction\cite{Wang2012}. 

Due to the coexistence of two types of spin-momentum locking---TSS and RSS, the surface spin texture is expected to be more complicated in our system. 
In Fig.~3(e-g), we show the spin projections of $S_{x}$, $S_{y}$, and $S_{z}$ for the I-terminated surface of Bi$_3$Te$_4$I at different constant energies.  In Fig.~3(e1) at energy 250 meV, there are three contours: inner, middle, and outer ones.  All of them have the largest spin distributions along $k_{x}$ direction while suppressed values along $k_{y}$ direction. The outer and inner ones are from the RSS states which have opposite spin polarizations, while the middle one is from the TSS.  When shifting the energy level below the RSS crossing point [Fig.~3(e2)], the inner spin contour will change its polarization and then switches place with the middle TSS contour and merges with the outer one [Fig.~3(e3)], leaving only two spin contours. Below the conduction band edge, only the TSS contour remains and its spin distribution changes sign when the energy level crosses the Dirac point [Fig.~3(e4)]. The spin projection of $S_{y}$ shown in Fig.~3(f) shows similar features. Combining results in Fig.~3(e) and 3(f), one can easily obtain the in-plane spin chiralities of TSS and RSS states. For TSS, the upper and lower Dirac cones have the left- and right-handed spin textures respectively.  For RSS states, the inner-branch has the left- (right-) handed spin textures above (below) the RSS crossing point, while the outer-branch always has the right-handed spin texture.  The $S_{z}$ projection shown in Fig.~3(g) represents the out-of-plane spin texture. It can be observed that the outer RSS branch has the largest out-of-plane spin component than TSS and the inner branch of RSS. For the Te-terminated surface, only the spin projections for the upper Dirac cone are shown [Fig.~3(h)-3(j)] because the lower Dirac cone is deeply buried in the bulk valence bands.  The spin textures of RSS states are identical to those of I-terminated surface, hence are not shown here.

Apart from the spin-momentum locking discussed above, recent theoretical and experimental works found that in topological insulator Bi$_{2}$Se$_{3}$ 
there was an additional locking between spin and orbital for TSS, leading to a spin-orbital texture\cite{Cao2013,Zhang2013,Xie2013}. Following the method of Zhang \emph{et al.}\cite{Zhang2013}, we calculate the in-plane ($p_{x}$ and $p_{y}$) and out-of-plane ($p_{z}$) orbital projections together with the orbital-selective spin textures for 
TSS of both I-terminated and Te-terminated surfaces of Bi$_3$Te$_4$I. First let's examine the I-terminated surface. As shown in Fig.~4, for the in-plane orbitals, the right-handed spin texture is coupled to tangential (radial) orbital texture on the upper (lower) Dirac cone. (tangential and radial in-plane orbitals refer to linear combinations of $p_x$ and $p_y$ orbitals with their orbital axis along tangential or radial directions\cite{Zhang2013}.) For the out-of-plane $p_z$ orbital, it is coupled to the left-(right-) handed spin texture on the upper (lower) Dirac cone. As for the Te-terminated surface, the orbital-selective spin textures are opposite to the I-terminated ones. 
We have therefore demonstrated that the interesting spin-orbital texture of TSS, initially observed in Bi$_{2}$Se$_{3}$\cite{Xie2013}, also appears in our inversion asymmetric heterostructure systems.

\section*{Discussion}
As we mentioned before, in several aspects, the BiTeI/Bi$_2$Te$_3$ heterostructure retains the properties of the bulk BiTeI. 
However, the bulk BiTeI is a trivial insulator. It can only be converted to a topological insulator under pressure. 
In contrast, for the system which we study, no external constraint is required. It is the intercalation by layers of Bi$_2$Te$_3$ that
changes the topological order of the structure. 
To better understand this result, we start from the orbital analysis of Bi$_2$Te$_3$ and BiTeI. Let's focus on the conduction band edge. The states there are mainly from Bi $p$ orbitals. For a QL of Bi$_2$Te$_3$, the $p$ orbitals from the two Bi atoms can be combined to form bonding and antibonding orbitals. When crystal field is taken into account, $p_z$ orbital is split from $p_{x,y}$ orbitals, with the bonding orbital $p_{z}^{+} = \frac{1}{\sqrt{2}}(p_{z}^{Bi_{2}}-p_{z}^{Bi_{3}})$ having the lower energy\cite{Zhang2009}, as shown in the left shadowed area of Fig.~5. Similar energy splitting due to crystal field also occurs for BiTeI\cite{Bahramy2011}. 

When the layers of two materials are stacked together to form a heterostructure, due to the overlap between orbitals, the levels with close energies and the same symmetry character are coupled and repel each other.  
Using Bi$_3$Te$_4$I as an example, on the right hand side of Fig.~2(a), we show the Bi$_1$-$p_z$ (Bi in TLs) and $p_{z}^{+}$ (Bi in QLs) orbital projected character of the low energy conduction bands. 
At $\Gamma$ point, the $p_{z}^{+}$ and Bi$_1$-$p_z$ orbital are mainly concentrated in the lowest and the second lowest conduction band respectively. As moving towards A point,  the coupling between $p_{z}^{+}$ and Bi$_1$-$p_z$ orbitals is getting stronger, which pushes the lowest conduction band downward. 
Therefore the intercalated Bi$_2$Te$_3$ not only provides the  $p_{z}^{+}$  orbital which has a lower energy than the Bi$_1$-$p_z$ orbital, but also through the coupling between the two orbitals it helps to further reduce the band gap ( from 1.2 eV in bulk BiTeI to 0.52 eV in Bi$_3$Te$_4$I ). Hence it becomes easier to achieve the band gap inversion when SOC is turned on.

This band gap reduction induced by intercalation also helps to explain the observed giant RSS. As been pointed out in the study of bulk
BiTeI\cite{Bahramy2011}, one of the several factors contributing to a large RSS is that the band gap should be small. Approximately, the RSS strength is inversely
proportional to the gap size. The reduced gap size of the heterostructure helps to realize a giant RSS. This point is reflected from the results in Table II. The two structures ( [(BiTeI)$_2$(Bi$_2$Te$_3$)$_1$]  and [(BiTeI)$_3$(Bi$_2$Te$_3$)$_1$] ) with RSS energy greater than that of bulk BiTeI indeed have very small band gaps.

To conclude, we have investigated the crystal structure and electronic properties of
the heterostructures [(BiTeI)$_m$(Bi$_2$Te$_3$)$_n$] by first principles calculations. We discover that these heterostructures 
host both strong topological insulating phase and giant RSS in their native states.
Moreover, both the $\mathbb{Z}_2$ invariants that characterize the topological properties and the Rashba energy that characterizes strength of RSS 
can be controlled by
varying the layer numbers $m$ and $n$. The surface electronic structure and spin texture present unique features of the coexistence of TSS and RSS,
and of strongly inversion asymmetric topological insulators. Our result indicates a rich physics in this class of heterostructures which provide an ideal platform for future development of spintronics and quantum computation devices.

\section*{Methods}
\subsection{Structure optimizations.}
The lattice parameters and the atomic positions of the heterostructures are optimized using the projector augmented wave method\cite{Blochl1994} implemented in VASP code\cite{Kresse1993,Kresse1996}. Perdew-Burke-Ernzerhof parametrization of the generalized gradient approximation (GGA-PBE) is used for the exchange correlation potential\cite{Perdew1996}.  Since the interaction between TLs and QLs is of the vdW-type similar to bulk BiTeI and Bi$_2$Te$_3$,  the vdW corrections by the approach of Dion \emph{et al.}\cite{Dion2004} (vdW-DF) as implemented in VASP is employed to further relax the out-of-plane lattice parameter $c$ and atomic positions\cite{klime2011}.  The plane wave energy cutoff is set to 300 eV, and the Brillouin zone is sampled by a $16\times 16\times 4$ mesh.
Both the crystal's shape and the atomic positions are relaxed until force on each ion becomes less than 0.01 eV$\cdot$\AA$^{-1}$. The phonon spectrum is calculated using the PHONOPY code\cite{Togo2008} through the DFPT approach\cite{Gonze1997}. 

\subsection{Bulk electronic structure calculations.}
The electronic ground-state calculations are performed using full-potential linearized augmented plane-wave (FP-LAPW) method\cite{Singh1994}, implemented in the package \textsc{WIEN2k}\cite{Blaha2001}.  The convergence criteria are more stringent than those of structural optimizations, with $K_\mathrm{max}R_\mathrm{MT}=8.0$ and a $20\times 20\times 4$ \textit{k}-mesh, where $R_\mathrm{MT}$ represents the smallest muffin-tin radius (2.5 Bohr for all atoms) and $K_\mathrm{max}$ is the maximum size of reciprocal-lattice vectors.  Spin-orbit coupling is included by a second-variational procedure\cite{Singh1994}. 

\subsection{Surface electronic structure calculations.}
Surface electronic structures are calculated using the combination of maximally localized wannier function (MLWF)\cite{Marzari1997,Souza2001} and surface Green's function methods\cite{Sancho1984,Sancho1985}.  The construction of MLWF using package \textsc{wannier90}\cite{Mostofi2008} is a non-self-consistent process on a $8\times 8\times 3$ \textit{k}-mesh with formerly converged self-consistent charge potential.  The surfaces with different atom-terminations are considered as semi-infinite systems, from which the iterative method\cite{Sancho1984,Sancho1985} is used to obtain the surface Green's function.  The LDOS is extracted from the imaginary part of the surface Green's function, which is used to analyze surface electronic structures.

\begin{addendum}
\item [Acknowledgments]
This work was supported by the MOST Project of China (Grants Nos.~2014CB920903, 2011CBA00100), SUTD-SRG-EPD-2013062, NSFC (Grant Nos.~11174337, 11225418, 11135001 and 11374033), and the Specialized Research Fund for the Doctoral Program of Higher Education of China (Grants No. 20121101110046).

\item [Author Contributions]
Y. Y. conceived the idea and designed the research.  J. J. Z. performed calculations. 
J. J. Z., W. F., Y. Z., S. A. Y. and Y. Y. contributed to the analysis and interpretation of the results and prepared the manuscript. 

\item [Competing Interests]
The authors declare no competing financial interests.

\item [Correspondence]
Correspondence and requests for materials should be addressed to Shengyuan A. Yang or Yugui Yao.

\end{addendum}

\clearpage
\newpage
\bigskip
\textbf{Figure 1.}
(a) Crystal structure of the Bi$_{3}$Te$_{4}$I superlattice consisting of one QL of Bi$_{2}$Te$_{3}$ and one TL of BiTeI along the crystallographic \textit{c}-axis.  There are eight nonequivalent atoms in one unit cell with the stacking order of Bi$_{1}$-I-Te$_{1}$-Bi$_{2}$-Te$_{2}$-Bi$_{3}$-Te$_{3}$-Te$_{4}$ from bottom to top, as labeled in the figure. (b) Hexagonal Brillouin zone with the high-symmetry points.  The projected  surface 2D Brillouin zones are also shown.   (c) The calculated phonon spectrum of Bi$_3$Te$_4$I with SOC included.

\bigskip
\textbf{Figure 2.}
The band structure of Bi$_3$Te$_4$I without (a) and with (b) SOC.   For (a),  the Bi$_1$-$p_z$ (Bi in TLs) and Bi$_{2,3}$-$p_{z}^{+}$ (Bi in QLs) orbital projected character of bands in the dashed box is shown on the right side. The size of circle denotes the weight of projection. (c) band dispersion in the vicinity of $\Gamma$ points showing the Rashba splitting. The parameters of Rashba splitting are also illustrated in the figure. (d) the local band gap  at $k_z$=0.5 plane  (around A points) and $k_z$=0.0 plane  (around $\Gamma$ point) as a function of the strength of SOC, where $\lambda_{SO}=1$ represents the intrinsic value of SOC. (e) surface  energy spectrum of side-cut surface ($ac$ plane) of Bi$_3$Te$_4$I. 

\bigskip
\textbf{Figure 3.}
The surface energy spectrums of I-terminated (a), Te-terminated (b) surfaces of Bi$_{3}$Te$_{4}$I; and I-terminated (c), Te-terminated (d) surfaces of Bi$_{4}$Te$_{5}$I$_{2}$.  The Fermi level is set to the bulk valence band maximum and dotted lines in (a) and (b) denote various constant energy levels where spin textures are imaged.  In (e)-(g) the spin-resolved constant energy contours are plotted for the I-terminated surface of Bi$_3$Te$_4$I, while those for the Te-terminated surface are shown in (h)-(j). The red and blue colors denote the positive and negative values of spin projections of $S_{x}$, $S_{y}$, and $S_{z}$, respectively.  The arrows label the spin polarizations of the outermost contour.

\bigskip
\textbf{Figure 4.}
 The $p_{x}$ and $p_{y}$ orbital-selective spin textures of Bi$_{3}$Te$_{4}$I at the upper (a) and lower (b) Dirac cone for I-terminated surface and (c) the upper  Dirac cone for Te-terminated surface, respectively.  the corresponding $p_{z}$  orbital-selective spin textures are shown in (d), (e), and (f).  More red and blue colors in (a)-(c) denote more $p_{x}$ and $p_{y}$ character respectively, while more red colors in (d)-(f) denote more $p_{z}$ character.  The surface states in the energy range of -0.1 eV $\sim$ 0.2 eV are used for the orbital projections.  The black arrows represent the total in-plane spin textures related to the corresponding orbitals.

\bigskip
\textbf{Figure 5.}
Schematic diagram of the evolution from atomic orbitals of Bi, Te and I into the states near the band edges of the BiTeI/Bi$_2$Te$_3$ heterostructure in the absence of SOC. The evolution stages are explained in the main text. 

\newpage

\begin{table}
   \caption{\label{bond} Lattice parameters and bond lengths of crystal structures optimized using GGA-PBE and vdW-DF methods
   (see Fig.~1(a) for the atomic labels)}
   \begin{tabular}{cccccc}
   \hline
   \multicolumn{3}{c}{\textrm{}} &
   \multicolumn{3}{c}{\textrm{Bond length (\AA)}} \\
   \multicolumn{1}{c}{\textrm{}} &
   \multicolumn{1}{c}{\textrm{a (\AA)}} &
   \multicolumn{1}{c}{\textrm{c (\AA)}} &
   \multicolumn{1}{c}{\textrm{Bi$_1$-I}} &
   \multicolumn{1}{c}{\textrm{I-Te$_1$}} &
   \multicolumn{1}{c}{\textrm{Te$_1$-Bi$_2$}} \\
   \hline
 \textrm{GGA-PBE} & 4.431 & 18.555 &  3.291 &  4.595 &  3.082  \\
 \textrm{vdW-DF}  & 4.431 & 16.921 &  3.295 &  3.850 &  3.083  \\
   \hline
   \multicolumn{2}{c}{\textrm{}} &
   \multicolumn{4}{c}{\textrm{}} \\
   \multicolumn{1}{c}{\textrm{}} &
   \multicolumn{1}{c}{\textrm{Bi$_2$-Te$_2$}} &
   \multicolumn{1}{c}{\textrm{Te$_2$-Bi$_3$}} &
   \multicolumn{1}{c}{\textrm{Bi$_3$-Te$_3$}} &
   \multicolumn{1}{c}{\textrm{Te$_3$-Te$_4$}} &
   \multicolumn{1}{c}{\textrm{Te$_4$-Bi$_1$}} \\
   \hline
   \textrm{GGA-PBE} & 3.292  & 3.278 &  3.087 &  4.249 &  3.075  \\
   \textrm{vdW-DF}  & 3.280  & 3.254 &  3.094 &  3.746 &  3.076  \\
   \hline
   \end{tabular}
\end{table}

\begin{table}
 \caption{\label{heterostructures} Energy gap ($E_g$ in meV), Rashba splitting energy ($E_R$ in meV)  and $\mathbb{Z}_2$ invariants in heterostructures with different layer configurations. 
 (TL$_m$QL$_n$ refers to the structure [(BiTeI)$_m$ (Bi$_2 $Te$_3 $)$_n$])}
 \begin{tabular}{ccccccc}
 \hline
 \multicolumn{1}{c}{\textrm{}} &
 \multicolumn{1}{c}{\textrm{TL$_1$QL$_1$}} &
 \multicolumn{1}{c}{\textrm{TL$_1$QL$_2$}} &
 \multicolumn{1}{c}{\textrm{TL$_1$QL$_3$}} &
 \multicolumn{1}{c}{\textrm{TL$_2$QL$_1$}} &
 \multicolumn{1}{c}{\textrm{TL$_2$QL$_2$}} &
 \multicolumn{1}{c}{\textrm{TL$_3$QL$_1$}} \\
  \hline
 $E_g$   & 117 & 114 &  84 &  22 &  52 & 14 \\
 \hline
 $E_R$   & 76 & 38 &  22  &  130 &  62 & 139 \\
 \hline
 $Z_2$   &  1;(001) & 1;(001) &  1;(001) &  1;(000) &  1;(000) & 1;(001) \\
 \hline
 \end{tabular}
\end{table}

\clearpage
\newpage

\begin{figure}
  \begin{center}
   \epsfig{file=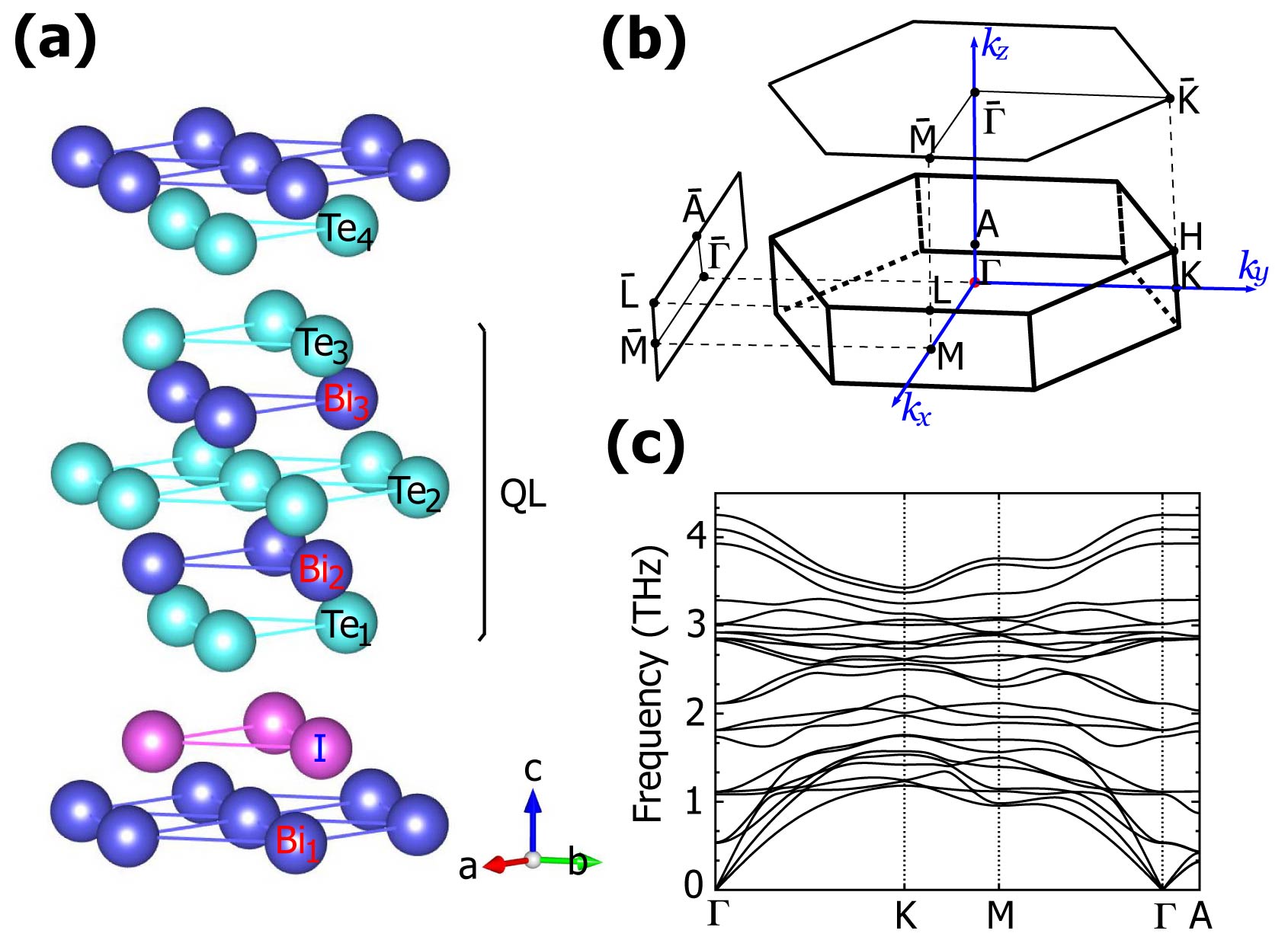,width=14cm}
  \end{center}
   \label{fig1} 
\end{figure}

\newpage

\begin{figure}
  \begin{center}
   \epsfig{file=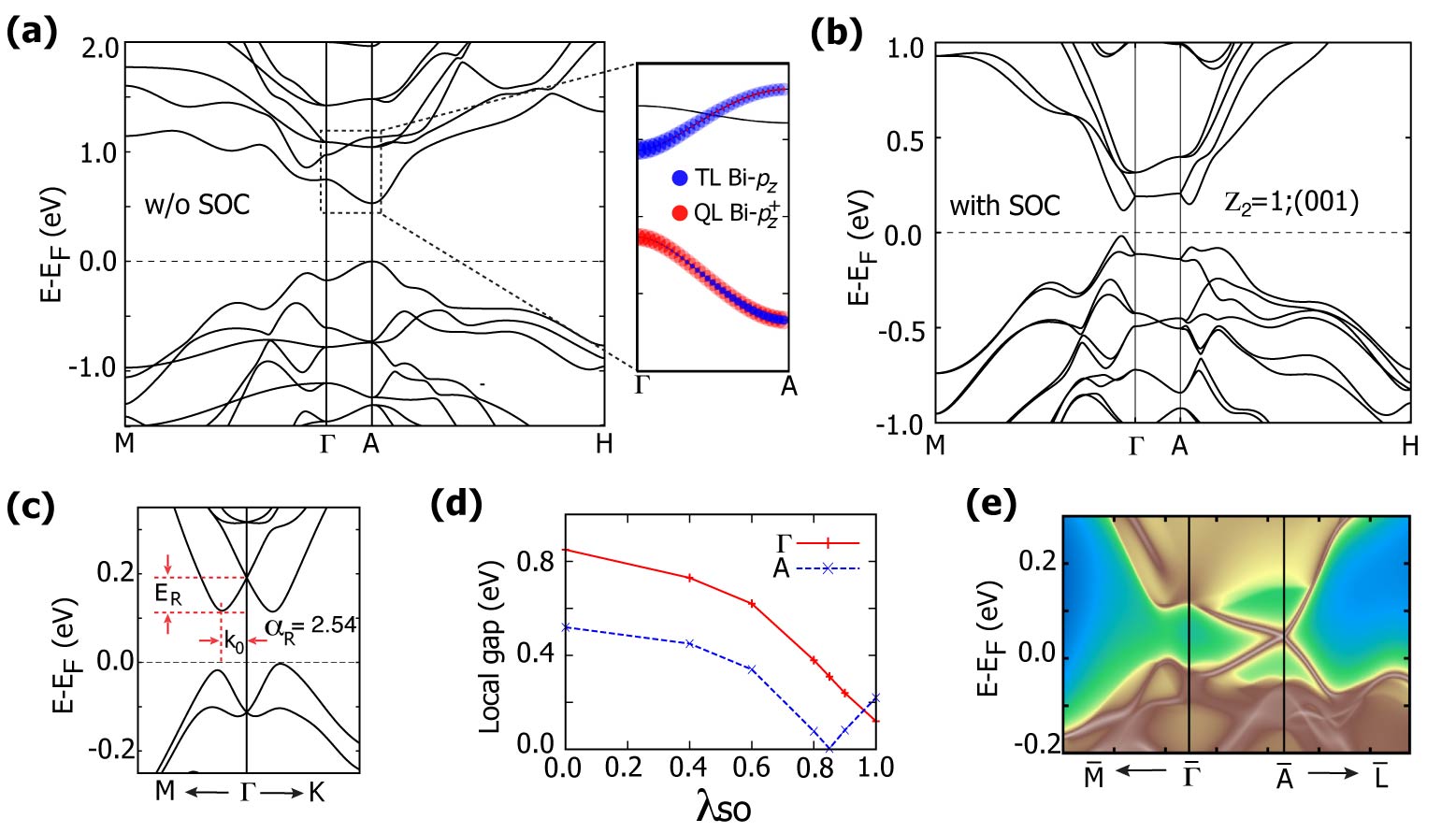,width=16cm}
  \end{center}
 \label{fig2} 
\end{figure}

\newpage

\begin{figure} 
  \begin{center}
   \epsfig{file=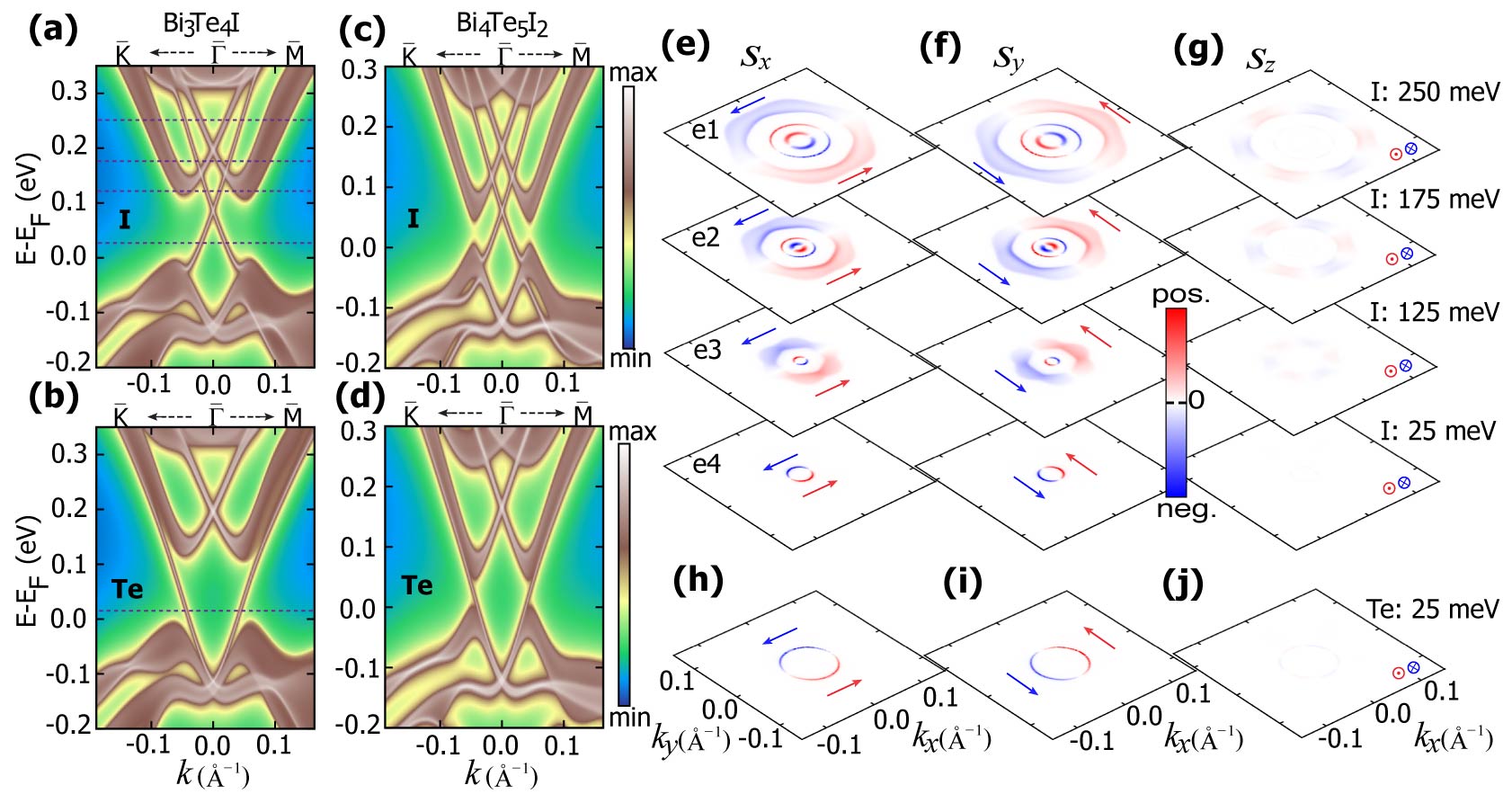,width=17cm}
  \end{center}
 \label{fig3} 
\end{figure}

\newpage

\begin{figure}
   \begin{center}
   \epsfig{file=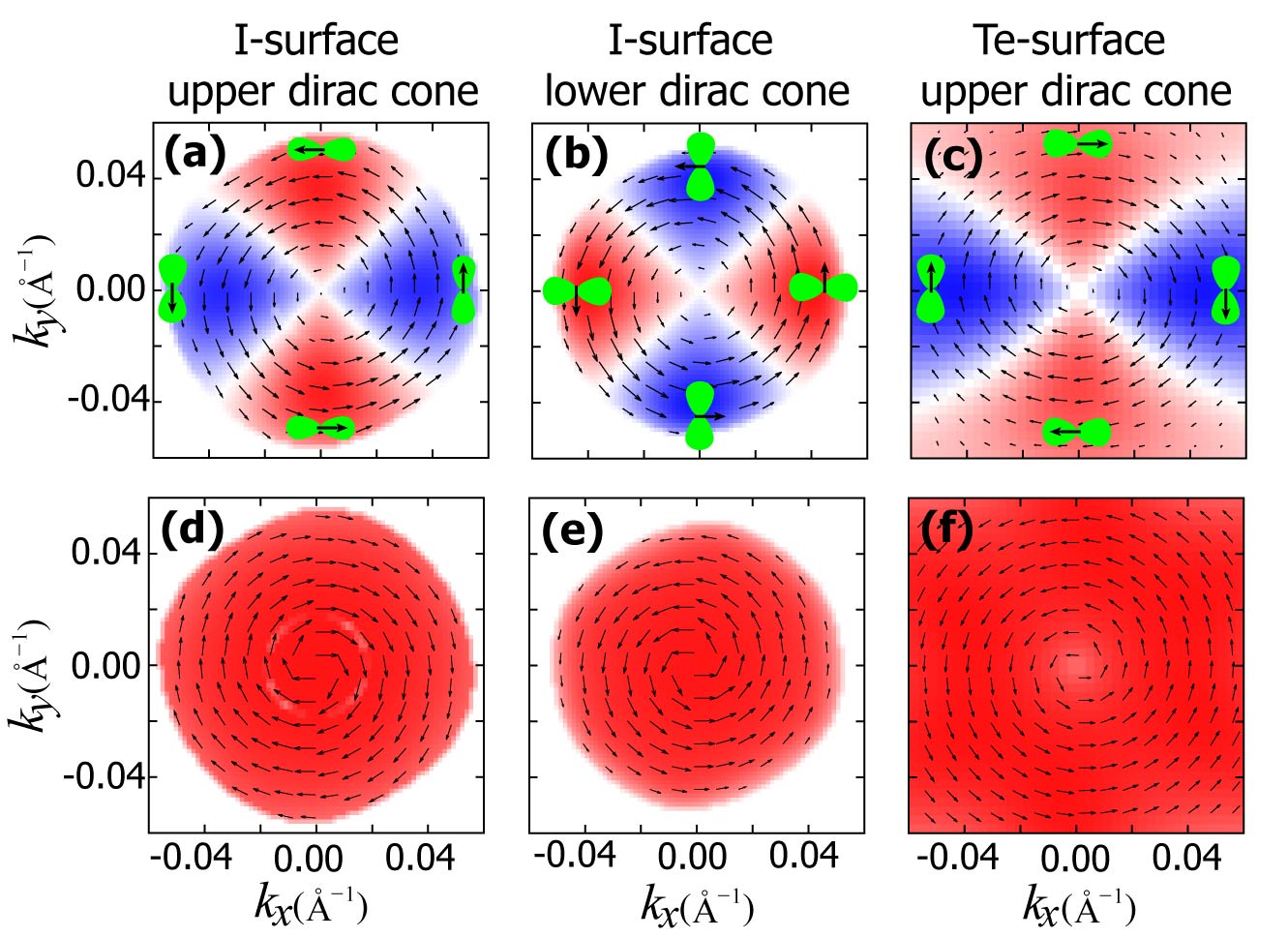,width=14cm}
  \end{center}
 \label{fig4}
\end{figure}

\newpage

\begin{figure}
  \begin{center}
   \epsfig{file=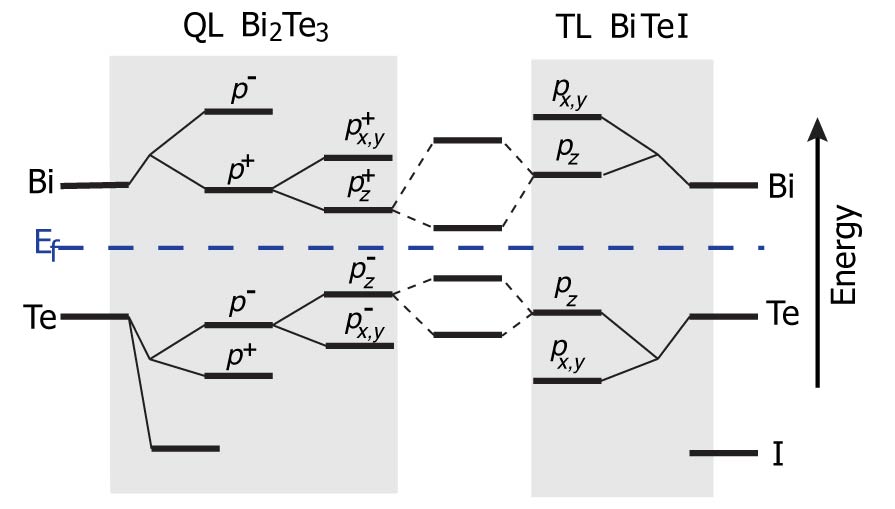,width=14cm}
  \end{center}
 \label{fig5} 
\end{figure}

\end{document}